\documentclass[usenatbib,referee]{mn2e}
\usepackage{graphicx}
\usepackage{epsfig}
\usepackage{color}

\title[He-accreting CO WDs and SNe Ia]
{He-accreting carbon-oxygen white dwarfs and type Ia supernovae}
\author[Wang, Podsiadlowski \& Han]
{Bo Wang,$^{\rm 1,2}$\thanks{E-mail:wangbo@ynao.ac.cn} Philipp Podsiadlowski,$^{\rm 3}$\thanks{E-mail:podsi@astro.ox.ac.uk} and Zhanwen Han$^{\rm 1,2}$\thanks{E-mail:zhanwenhan@ynao.ac.cn} \\
$^1$Key Laboratory for the Structure and Evolution of Celestial Objects, Yunnan Observatories, CAS,
Kunming 650216, China\\
$^2$Center for Astronomical Mega-Science, CAS, Beijing 100012, China\\
$^3$Department of Astronomy, Oxford University, Oxford OX1 3RH, UK}

\begin{document}
\date{Accepted. Received}
\pagerange{\pageref{firstpage}--\pageref{lastpage}} \pubyear{2017}
\maketitle

\label{firstpage}

\begin{abstract}
He accretion onto carbon-oxygen white dwarfs (CO WDs) plays a
fundamental role when studying the formation of type Ia supernovae
(SNe Ia).  Employing the MESA stellar evolution code, we calculated
the long-term evolution of He-accreting CO WDs.  Previous studies
usually supposed that a WD can grow in mass to the Chandrasekhar limit
in the stable He burning region and finally produce a SN Ia.  However,
in this study we find that off-centre carbon ignition occurs in the
stable He burning region if the accretion rate is above a critical
value ($\sim$$2.05\times10^{-6}\,{M}_{\odot}\,\mbox{yr}^{-1}$), resulting
in accretion-induced collapse rather than a SN Ia. If the accretion
rate is below the critical value, explosive carbon ignition will
eventually happen in the centre producing a SN Ia.  Taking into
account the possibility of off-centre carbon ignition, we have
re-determined the initial parameter space that produces SNe Ia in the
He star donor channel, one of the promising channels to produce SNe Ia
in young populations. Since this parameter space is smaller than was
found in the previous study of Wang et al.\ (2009), the SN Ia rates
are also correspondingly smaller.  We also determined the chemical
abundance profile of the He-accreting WDs at the moment of explosive
carbon ignition, which can be used as initial input for SN Ia
explosion models.
\end{abstract}

\begin{keywords}
binaries: close --  stars: evolution  -- white dwarfs -- supernovae: general
\end{keywords}

\section{Introduction}

Type Ia supernovae (SNe Ia) play an important role in cosmology and
in our current understanding of the chemical evolution of galaxies
(e.g. Matteucci \& Greggio 1986; Howell 2011; Meng, Gao \& Han 2015).
There is some consensus that SNe Ia arise from thermonuclear explosions
of carbon-oxygen white dwarfs (CO WDs) in binaries, although the mass
donor is still uncertain (e.g. Hoyle \& Fowler 1960; Nomoto, Iwamoto
\& Kishimoto 1997; Podsiadlowski et al.\ 2008).  The mass donor could
be a non-degenerate star in the single-degenerate (SD) model or
another WD in the double-degenerate (DD) model (e.g. Whelan \& Iben
1973; Webbink 1984; Iben \& Tutukov 1984).  For recent reviews on the
progenitor issue of SNe Ia see, e.g.  Wang \& Han (2012),
H\"{o}flich et al. (2013), Hillebrandt et al. (2013), Ruiz-Lapuente
(2014), and Maoz, Mannucci \& Nelemans (2014).

In the classical SD model, a WD accretes H-rich matter from a MS star
or a RG star (e.g. Hachisu, Kato \& Nomoto 1996; Li \& van den Heuvel 1997; Langer et al. 2000;
Han \& Podsiadlowski 2004, 2006; Wang, Li \& Han 2010; Meng \& Podsiadlowski 2017). It is also possible
that a WD accretes He-rich matter from a He star or a He subgiant to
grow in mass to the Chandrasekhar limit and then produce a SN Ia,
which is referred to as the He star donor channel (see Yoon \& Langer
2003; Wang et al. 2009a,b).  It has been suggested that the He star
donor channel is a particularly favourable channel for producing
observed SNe Ia with short delay times (e.g.  Mannucci, Della Valle \&
Panagia 2006; Cooper, Newman \& Yan 2009; Wang et al. 2009a,b; Ruiter
et al. 2009; Thomson \& Chary 2011).  Many recent binary population
synthesis (BPS) studies also involved the He star donor channel
(e.g. Ruiter et al. 2013, 2014; Toonen et al. 2014; Claeys et
al. 2014).

Observationally, many WD+He star systems have been considered as
progenitor candidates for SNe Ia, e.g.  V445 Puppis (see Kato et
al. 2008; Woudt et al. 2009), HD 49798 with its WD companion (see Wang
\& Han 2010a; Mereghetti et al. 2011; Liu et al. 2015),
CD$-$30$^{\circ}$\,11223 (see Vennes et al. 2012; Geier et al. 2013;
Wang, Justham \& Han 2013), and KPD 1930+2752 (see Maxted, Marsh \&
North 2000; Geier et al. 2007), etc.  Especially, V445 Puppis is a
strong candidate for a SN Ia progenitor as the WD mass is at least
1.35\,${M}_\odot$ and the mass retention efficiency of He accretion
onto the WD appears to be as high as 50\,\% during nova outbursts
(e.g. Kato et al. 2008).  In addition, SN 2014J may originate from a
WD+He star system, and the mass donor for the progenitor of SN 2012Z
may have been a He star (e.g. McCully et al. 2014; Diehl et al. 2014;
Wang et al. 2014a).  Moreover, the hypervelocity He star US 708 and
its spectroscopic twin J2050 could be surviving mass donors of SNe Ia
that occurred in WD+He star systems (e.g. Justham et al. 2009; Wang \&
Han 2009; Geier et al. 2015; Ziegerer et al. 2017).  Note that WD+He
star systems are also involved in the formation of some peculiar types
of systems, such as AM CVn binaries (e.g. Nelemans et al. 2001; Brooks
et al. 2015; Piersanti, Tornamb\'{e} \& Yungelson 2015) and double CO WDs
(e.g. Ruiter et al. 2013; Liu et al. 2016).

However, the accretion of He-rich matter onto WDs is still not
completely understood.  Previous studies indicated that the accretion
rate plays a crucial role in the evolution of the He-accreting WDs
(e.g.  Kato \& Hachisu 2004; Piersanti, Tornamb\'{e} \& Yungelson
2014; Wang et al. 2015; Wu et al. 2016).  If the accretion rate is too high, the WD
will develop into a red-giant-like He star due to the continuous
pileup of the accreted He-rich matter on its surface; if the accretion
rate is too low, it will experience He-shell flashes due to unstable
nuclear burning (e.g. Nomoto 1982a; Kato \& Hachisu 2004; Wu et al. 2017).  Previous
studies usually assumed that a WD can grow steadily in mass to the
Chandrasekhar limit in a narrow parameter region for stable He burning
and then produce a SN Ia (e.g. Nomoto 1982a; Wang et al. 2009a).  In
contrast, in the present work where we followed the long-term
evolution of the accreting WDs, we found that they can experience
either centre or off-centre carbon ignition when their mass is close
to the Chandrasekhar limit. Off-centre carbon burning will convert CO
WDs into ONe WDs, ultimately leading to the formation of neutron stars
rather than SNe Ia (e.g. Saio \& Nomoto 1985, 1998; Nomoto \& Iben
1985; Schwab, Quataert \& Kasen 2016).  This should affect the SN Ia
rates in the He star donor channel, which will be studied in this
work.  Note that Brooks et al. (2016) recently also reported
  these two possible outcomes, but they only considered a narrow
  binary parameter space.  Here, we explore this question in a
  systematic way and apply the results using a detailed BPS approach.

In most previous studies of the SD model, the WDs are taken as point
masses, and their structure is not calculated when simulating mass
accretion; the WDs are supposed to explode as SNe Ia once they have
grown in mass to the Chandrasekhar limit (e.g. L\"{u} et al.  2009;
Toonen et al. 2012; Meng \& Yang 2012; Bours, Toonen
\& Nelemans 2013). In this study, however, we calculate the
long-term evolution of the He-accreting CO WDs by solving their
structure equations. In Section 2, we introduce the basic assumptions
and methods for the numerical simulations.  The numerical results of our
simulations are given in Section 3.  In Section 4, we show the initial
parameter space for SNe Ia based on the He star donor channel and the
corresponding BPS results.  Finally, we discuss the results in Section
5 and give a summary in Section 6.

\section{Numerical method}

\subsection{Stellar evolution code}

Using the MESA stellar evolution code (version 7624; see
Paxton et al. 2011, 2013, 2015), we calculate the long-term evolution
of He-accreting CO WDs.  The default OPAL opacity is used in our
simulations, and the nuclear reaction network \texttt{co\_burn.net} is
adopted.  This nuclear reaction network contains isotopes needed for
helium, carbon and oxygen burning, which are coupled by 57 nuclear
reactions.  Here, we adopt two established cases in MESA
(\texttt{make\_co\_wd} and \texttt{wd2}) to perform our simulations.
The established case \texttt{make\_co\_wd} is used to build initial
models of CO WDs, whereas the established case \texttt{wd2} is used to
simulate the long-term evolution of He-accreting CO WDs.  The
established case \texttt{wd2} contains an acceleration term in the
hydrostatic equilibrium equation so that we can also simulate
He-shell flashes.

The initial WD models in our simulations have the following
  masses (temperatures) in units of ${M}_\odot$ ($10^7{\rm K}$): 0.6
  (7.1), 0.7 (7.0) , 0.8 (6.9), 0.9 (7.7), 1.0 (10.0), 1.1 (10.4), 1.2
  (12.2), 1.25 (13.7), 1.3 (16.0), and 1.35 (20.7).  The primordial
  metallicity for these WDs was taken as 0.02.  We performed a large
  number of calculations of He accretion onto the WDs varying the
  accretion rates in the range of $\dot{M}_{\rm
    acc}=10^{-8}-10^{-5}\,M_\odot\,\mbox{yr}^{-1}$ with a step size
  of $\delta\dot{M}_{\rm
    acc}=5\times10^{-8}\,{M}_\odot\,\mbox{yr}^{-1}$.  The accreted
He-rich matter consists of 98\,\% He and has a metallicity of 2\,\%.
In our calculations, the WDs were resolved with more than 2000 meshpoints.

\subsection{Criteria for explosive carbon ignition}

The accreted He can be transformed into carbon and oxygen completely
if the steady He-shell burning happens on the surface of the accreting
WD, increasing the mass of the WD in the process.  When the WD mass
approaches the Chandrasekhar limit, carbon in its centre will be ignited, which
may lead to explosive carbon burning.  At that moment, a huge amount
of thermonuclear energy is released by the explosive carbon burning
and produces a thermonuclear runaway if the energy generated cannot be
transported away by convection.  The WD will be destroyed by the
explosive carbon burning, leading to a SN Ia explosion.

Previous studies supposed that SNe Ia occur when the WDs evolve to
the point $t_{\rm b}={1}/{22}t_{\rm c}$ (e.g. Lesaffre et al. 2006),
where $t_{\rm c}$ is the timescale it takes a convective element to
cross a pressure scaleheight and $t_{\rm b}$ is the exponential
temperature growth time caused by the carbon burning.  However, the
specific point of explosive carbon ignition is still unclear
(e.g. Lesaffre et al. 2006; Chen, Han \& Meng 2014).  In the present
work, we found that the central density of the WD stops changing but
that the temperature increases dramatically after the He-accreting WD
reaches the point $t_{\rm b}={1}/{22}t_{\rm c}$ in Lesaffre et
al. (2006).  Similarly to the previous work of Chen, Han \& Meng
(2014), we take the point when the temperature starts to increase
  sharply in the centre as the point of the explosive carbon
ignition.

\section{Numerical Results}

\subsection{Properties of the He-shell burning}

\begin{figure}
\begin{center}
\epsfig{file=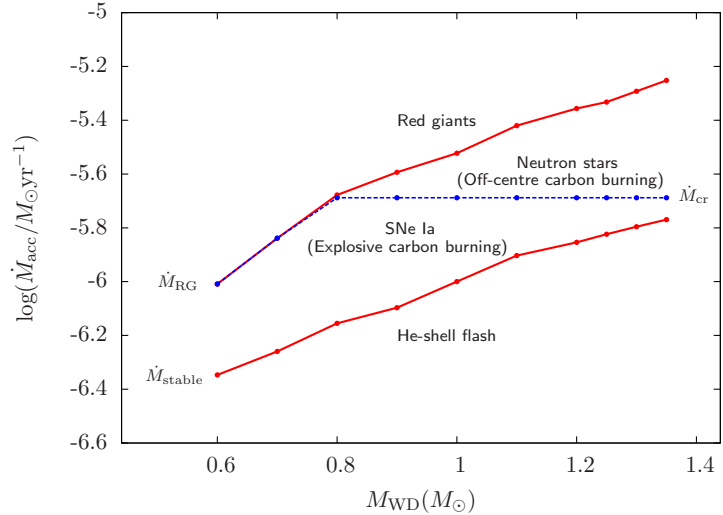,width=11cm} \caption{Stable He-shell burning
  region in the ${M}_{\rm WD}-\dot{M}_{\rm acc}$ plane.  The dotted
  line gives the critical accretion rate above which off-centre carbon
  ignition occurs when the WD mass approaches the Chandrasekhar limit.}
\end{center}
\end{figure}

We carried out a series of calculations with initial WD masses
of ${M}^{\rm i}_{\rm WD}=0.6-1.35\,{M}_\odot$ and accretion rates
of $\dot{M}_{\rm acc}=10^{-8}-10^{-5}\,M_\odot\,\mbox{yr}^{-1}$.  In
Fig. 1, we show the stable He-shell burning region in the ${M}_{\rm
  WD}-\dot{M}_{\rm acc}$ plane, in which the WD can grow steadily in mass.
If $\dot{M}_{\rm acc}$ is larger than the maximum
accretion rate $\dot{M}_{\rm RG}$ for stable He-shell burning, the
envelope of the WD will expand to red-giant size.  If $\dot{M}_{\rm
  acc}$ is below the minimum accretion rate $\dot{M}_{\rm stable}$ for
stable He-shell burning, the accreting WD will experience He-shell
flashes.  The values of $\dot{M}_{\rm stable}$ and
$\dot{M}_{\rm RG}$ (in $M_\odot\,\mbox{yr}^{-1}$) can be fitted with the
following formulae
\begin{equation}
\small{\dot{M}_{\rm stable}=1.46\times10^{-6}(-{M}_{\rm WD}^3+3.45{M}_{\rm WD}^2-2.60{M}_{\rm WD}+0.85)},
\end{equation}
\begin{equation}
\small{\dot{M}_{\rm RG}=2.17\times10^{-6}({M}_{\rm WD}^2+0.82{M}_{\rm WD}-0.38)},
\end{equation}
where ${M}_{\rm WD}$ is in units of $M_\odot$.
These two fits were obtained using a bisection method for different WD
masses and  accretion rates.

It has been supposed that a WD can grow in mass to the Chandrasekhar
limit in the stable He-shell burning region and then explode as a SN
Ia (e.g. Nomoto 1982a; Wang et al. 2009a).  However, in the present
work we found that off-centre carbon ignition occurs if $\dot{M}_{\rm
  acc}$ is above a critical value $\dot{M}_{\rm cr}$
($\sim$$2.05\times 10^{-6}\,{M}_\odot\,\mbox{yr}^{-1}$).  Off-centre
carbon ignition will convert CO WDs into ONe WDs via an
inward-propagating carbon flame, which will ultimately lead to the
formation of neutron stars through accretion induced collapse rather
than thermonuclear explosions (e.g.\ Saio \& Nomoto 1985, 1998; Nomoto
\& Iben 1985; Brooks et al. 2016; Schwab, Quataert \& Kasen 2016).
The WD can increase its mass steadily in the region between
$\dot{M}_{\rm stable}$ and $\dot{M}_{\rm cr}$, in which explosive
carbon ignition can happen in the centre of the WD, resulting in a SN
Ia explosion.

\subsection{Centre and off-centre carbon ignition}

\begin{figure}
\begin{center}
\epsfig{file=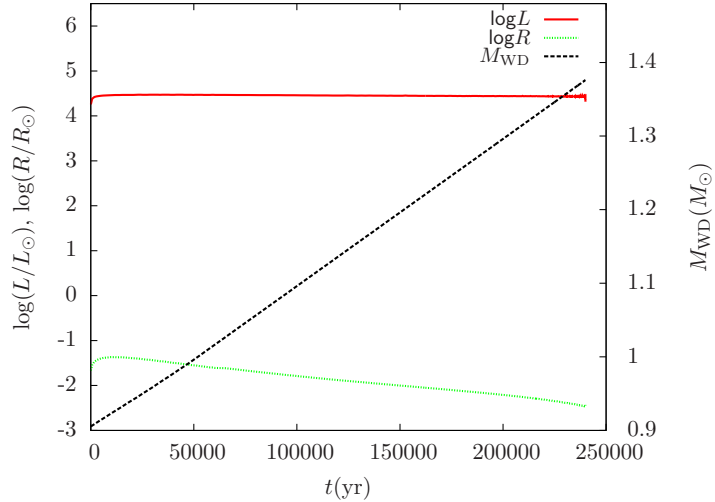, width=11.3cm}
\caption{An example of central carbon ignition, in which ${M}^{\rm i}_{\rm WD}=0.9\,{M}_\odot$ and
$\dot{M}_{\rm acc}=2\times10^{-6}\,{M}_\odot\,\mbox{yr}^{-1}$. The long-term evolution of the mass,  luminosity
and radius of the He-accreting WD are presented.}
\end{center}
\end{figure}

In Figs\ 2--5 we present the results of a representative example of
central carbon ignition, where ${M}^{\rm i}_{\rm WD}=0.9\,{M}_\odot$
and $\dot{M}_{\rm acc}=2\times10^{-6}\,{M}_\odot\,\mbox{yr}^{-1}$.
Fig.\ 2 shows the long-term evolution of the mass, luminosity and
radius of the He-accreting WD, where the initial central temperature
(${T}_{\rm c}$) and density (${\rho}_{\rm c}$) are $7.7\times10^7{\rm
  K}$ and $1.757\times10^7{\rm g}\,\mbox{cm}^{-3}$, respectively.
This figure shows that the mass of the WD grows linearly with
  time, as assumed in the model, that the WD radiates at the
  luminosity corresponding to steady He burning and that the radius of
  the WD follows the expected mass--radius relation.  He-shell
burning transforms the He-rich matter into carbon and oxygen,
increasing the mass of the CO core as a consequence.  The He-accreting
WD can grow steadily in mass until it approaches a mass of
$1.376\,{M}_\odot$; this phase lasts about
$2.35\times10^{5}\,\mbox{yr}$.  It is worth noting that the WD mass at
the time of explosion may exceed the Chandrasekhar limit if rotation is taken
into account (e.g. Yoon \& Langer 2004; Chen \& Li 2009; Justham 2011;
Hachisu et al. 2012; Wang et al. 2014b).

\begin{figure}
\begin{center}
\epsfig{file=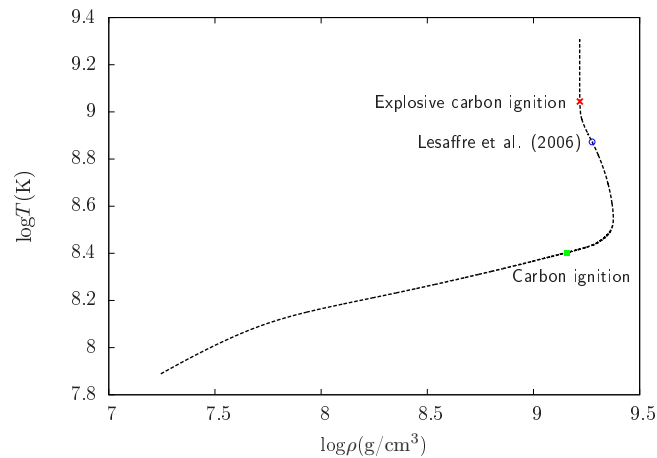,width=9.2cm} \caption{Evolution of ${\rho}_{\rm c}$
  and ${T}_{\rm c}$ for the He-accreting WD.  The red cross shows the
  starting point for explosive carbon burning that we used, whereas
  the blue open circle represents the point $t_{\rm b}={1}/{22}t_{\rm c}$
  in Lesaffre et al. (2006).  The green filled square indicates the starting
  point for carbon burning.}
\end{center}
\end{figure}

In Fig.\,3, we show the whole evolution of ${\rho}_{\rm c}$ and
${T}_{\rm c}$ of the He-accreting WD.  After about
$2.27\times10^{5}\,\mbox{yr}$, carbon is ignited in the centre of the
He-accreting WD, but initially is non-explosive as the thermonuclear
energy is transported away by convection.  At the end of the
simulation, ${\rho}_{\rm c}$ no longer evolves but the nuclear
reaction rate of carbon burning in the centre of the WD increases
quickly with ${T}_{\rm c}$,  which we identify with the start
  of {\em explosive} carbon burning.  The WD in our simulations can
grow in mass to the condition of explosive carbon ignition, resulting
in a SN Ia explosion.  It takes about $8\times10^{3}\,\mbox{yr}$ from
the point of carbon ignition to explosive carbon burning.

\begin{figure}
\begin{center}
\epsfig{file=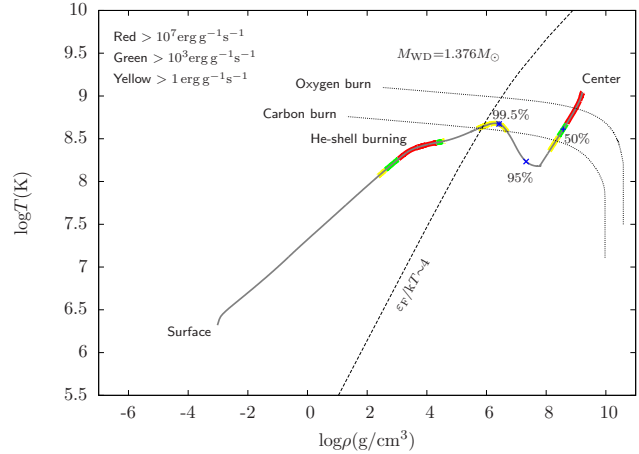,width=9cm} \caption{Profile of ${\rho}$ and ${T}$
  for a $1.376\,{M}_\odot$ WD at the point of explosive carbon
  ignition.  Different mass fractions inside the WD (50\%, 95\% and
  99.5\%) are indicated on the profile.  Degenerate and non-degenerate
  regions are separated by the dashed curve ($\varepsilon_{\rm F}/{\rm
    k}{T} \sim 4$), whereas the dotted curves represent the carbon and
  oxygen burning ignition curves, respectively. }
\end{center}
\end{figure}

\begin{figure}
\begin{center}
\epsfig{file=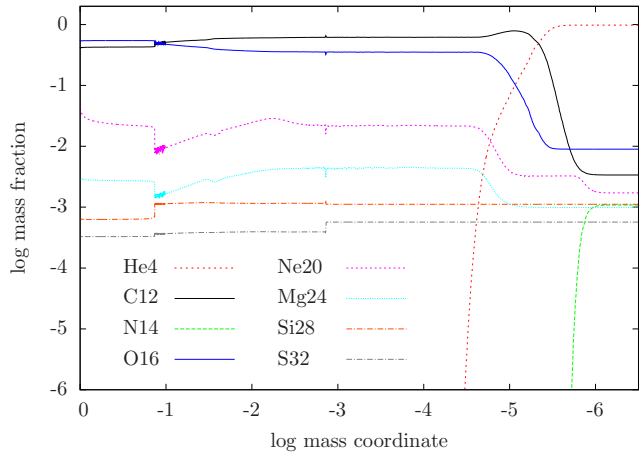,angle=0,width=9.7cm}
\caption{Chemical abundance profile at the point of explosive carbon ignition.}
\end{center}
\end{figure}

Fig.\,4 shows the ${\rho}-{T}$ profile of the WD when explosive carbon
ignition occurs, where ${\rho}$ and ${T}$ reach maximum
values of $1.65\times10^{9}{\rm g}\,\mbox{cm}^{-3}$ and
$1.11\times10^{9}{\rm K}$, respectively.  At this moment, a
deflagration wave starts to spread from the centre of the WD.  In
Fig. 5, we present the profile of some key chemical abundances at the
moment of explosive carbon ignition for this particular example.  The
accreted He is burnt into carbon, oxygen and other intermediate-mass
elements in the outer layer of the WD core.  This chemical abundance
profile can be taken as initial input for SN Ia explosion models.
The abundance profile (together with density, temperature,
  etc.) varies for different initial models and is made available on
  request by contacting BW.

\begin{figure}
\begin{center}
\epsfig{file=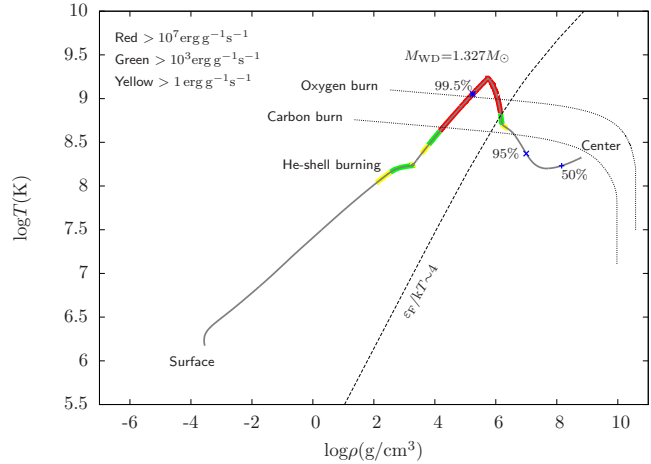,width=9.2cm}
\caption{Similar to Fig. 4, but for the ${\rho}-{T}$ profile of a
  $1.327\,{M}_\odot$ WD at the moment of off-centre carbon ignition,
  where ${M}^{\rm i}_{\rm WD}=1.0\,{M}_\odot$ and $\dot{M}_{\rm
    acc}=3\times 10^{-6}\,{M}_\odot\,\mbox{yr}^{-1}$}.
\end{center}
\end{figure}

If $\dot{M}_{\rm acc}$ is larger than $\dot{M}_{\rm cr}$, the
He-accreting WD will experience off-centre carbon ignition, similar to
what happens during the merging of double CO WDs (for more discussion
see Sect. 5).  In Fig. 6, we present a representative example of
off-centre carbon ignition, where ${M}^{\rm i}_{\rm
  WD}=1.0\,{M}_\odot$ and $\dot{M}_{\rm acc}=3\times
10^{-6}\,{M}_\odot\,\mbox{yr}^{-1}$.  In this case, it is the
  compressional heating of the outer layers caused by fast accretion
  that leads to off-centre ignition.  Off-centre carbon ignition
occurs in this example when the WD reaches a mass of
$1.327\,{M}_\odot$, while He-shell burning continues on the surface of
the WD.  It takes $1.08\times10^{5}\,\mbox{yr}$ from the beginning of
mass accretion to the condition of off-centre carbon ignition.  The
resulting carbon burning front likely propagates inwards in a quiet
manner, forming first an ONe WD but ultimately collapsing to form a
neutron star rather than a SN Ia (e.g.  Saio \& Nomoto 1985, 1998;
Nomoto \& Iben 1985; Brooks et al. 2016).

\section{The He star donor channel}

\subsection{Initial parameter space for SNe Ia}

Wang et al. (2009a) carried out a systematic study of the He star
donor channel for the progenitors of SNe Ia.  They performed binary
evolution calculations of the donor star with the Cambridge
stellar evolution code (Eggleton 1973; Han, Podsiadlowski \& Eggleton
1994; Pols et al.\ 1998) to determine the initial parameter space of
WD binaries that can result in SNe Ia in the orbital period--secondary
mass ($\log P^{\rm i}-M^{\rm i}_2$) plane.  Using these results and
adopting a detailed BPS approach, Wang et al. (2009b) then obtained
the SN Ia rates and delay times for the He star donor channel.
However, Wang et al. (2009a) did not take into account the possibility
of off-centre carbon ignition, which will reduce the
initial parameter space for SNe Ia and decrease the theoretical
SN Ia rates.  In the present work, we extract the mass-transfer
  rate from the data files of the binary evolution calculations in
  Wang et al. (2009a) and assume that off-centre carbon burning
  happens if the mass-transfer rate is higher than the critical
  value $\dot{M}_{\rm cr}$ in Sect. 3.1 when the CO WD grows in mass
  close to the Chandrasekharf limit.

\begin{figure}
\begin{center}
\epsfig{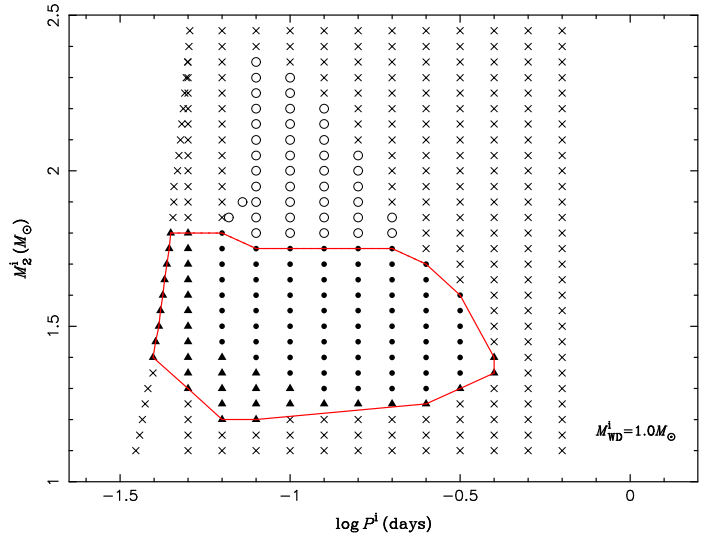} \caption{Initial parameter
  space for SNe Ia in the $\log P^{\rm i}-M^{\rm i}_2$ plane for
  ${M}^{\rm i}_{\rm WD}=1.0\,M_{\odot}$.  The filled symbols represent
  systems that lead to SN Ia explosions, where the filled triangles
  and circles indicate that the WDs explode as SNe Ia in the weak
  He-shell flash stage or in the stable He-shell burning stage,
  respectively.  Open circles show systems that experience off-centre carbon
  ignition, resulting in the eventual formation of neutron stars.
  Crosses denote systems that experience strong He-shell flashes
  which prevent WDs from growing in mass to the Chandrasekhar limit.}
\end{center}
\end{figure}

Fig. 7 shows the final outcomes of the binary evolution
calculations in the $\log P^{\rm i}-M^{\rm i}_2$ plane for ${M}^{\rm
  i}_{\rm WD}=1.0\,M_{\odot}$, where the filled symbols indicate
systems that result in SN Ia explosions.  The filled triangles and
circles in this figure denote that the WDs explode as SNe Ia in the
weak He-shell flash stage and in the stable He-shell burning stage,
respectively.  Some systems fail to form SNe Ia because strong
He-shell flashes prevent the WDs from growing in mass to the Chandrasekhar limit
(the crosses in Fig. 7). The open circles indicate systems that
experience off-centre carbon ignition and ultimately produce neutron
stars; these had been assumed to produce SNe Ia in Wang et
al. (2009a).

\begin{figure}
\begin{center}
\epsfig{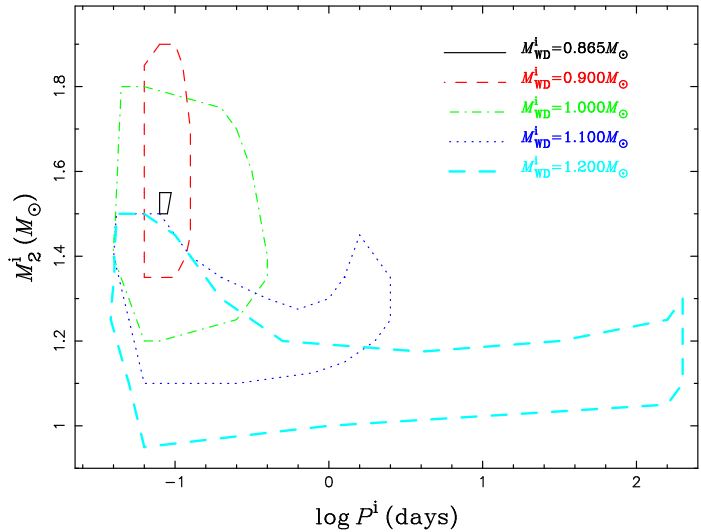} \caption{Initial parameter
  space for SNe Ia in the $\log P^{\rm i}-M^{\rm i}_2$ plane for
  different values of ${M}^{\rm i}_{\rm WD}$. }
\end{center}
\end{figure}

In Fig. 8, we present the initial parameter space for SNe Ia in the
$\log P^{\rm i}-M^{\rm i}_2$ plane for different values of ${M}^{\rm
  i}_{\rm WD}$.  For $M_{\rm WD}^{\rm i}=0.865\,M_{\odot}$, which is
close to the minimum WD mass for producing SNe Ia in the He star donor
channel, the possible parameter space is quite small.  The upper
boundaries of the parameter space are mainly constrained by the
condition for off-centre carbon ignition due to a high mass-transfer
rate when the WDs grow in mass close to the Chandrasekhar limit, which is
strongly dependent on the initial mass of the WD. The lower boundaries
are set by the condition that the mass-transfer rate should be high
enough to ensure that the WD can increase its mass to the Chandrasekhar limit.
The left boundaries are determined by the minimum value of $\log
P^{\rm i}$, for which a zero-age He MS star would fill its Roche
lobe.  WD+He star systems beyond the right boundaries experience a very
high mass-transfer rate because of the rapid expansion of He stars
during the subgiant phase; this drastically reduces the donor-star mass
through an optically thick wind (e.g. Hachisu, Kato \& Nomoto 1996); some of
these systems may contribute to the formation of double CO WDs and
produce SNe Ia through the DD model (e.g. Ruiter et al. 2013; Liu et
al. 2016).

The results obtained here are very similar to those of Brooks
  et al. (2016), as can be seen, e.g., by comparing these to Fig. 3
  from their paper. The main difference is that Brooks et al. (2016)
  have computed full binary evolution calculations of CO WD+He star
  systems and thus computed the (time-varying) mass accretion rates
  onto the WD self-consistently instead of assuming constant accretion
  rates. The resulting limits on the He donor mass leading to SNe Ia
  (about 1.3$-$1.7$\,M_{\odot}$) are nevertheless very similar in the
  two methods, at least for an initial WD mass of 1.0$\,M_{\odot}$.

\subsection{BPS results}

To obtain SN Ia rates for the He star donor channel, we carried
out a series of Monte Carlo BPS calculations based on the Hurley
binary evolution code (see Hurley, Tout \& Pols 2002).  The basic BPS
setup and principal assumptions here are similar to those of Wang et
al. (2009b),  but in the present work we used the updated
  initial parameter space for SNe Ia presented in Fig. 8.  In each
BPS simulation, the evolution of $4\times10^{\rm 7}$ sample binaries
is tracked from the primordial binary stage to the production of the
WD+He star systems (see Wang et al. 2009b).  We suppose that, if the
initial parameters of a WD+He star system are located inside the SN Ia
parameter space of Fig. 8, a SN Ia explosion happens.  To obtain the
outcome of the common-envelope ejection, the commonly used energy
equation of Webbink (1984) is employed.  As in previous
calculations, a free parameter $\alpha_{\rm ce}\lambda$ is used to
calculate the process of the common-envelope evolution, and its value
is set to 0.5 or 1.5 (e.g. Wang et al. 2009b).

\begin{figure}
\begin{center}
\epsfig{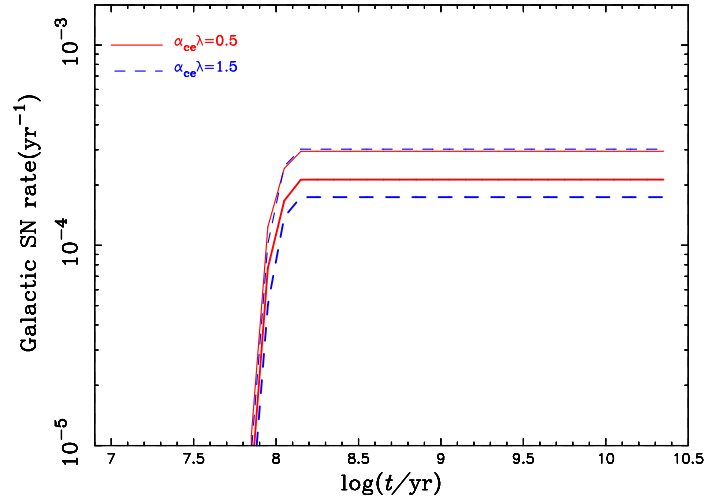}
\caption{Rates of SNe Ia in the Galaxy for a constant star-formation rate of $5\,M_{\odot}$\,yr$^{-1}$.  The thick curves
  are the results based on the initial contours in Fig. 8, whereas the
  thin curves are the results from Wang et al. (2009a). Different line
  styles are for different values of $\alpha_{\rm ce}\lambda$ as
  indicated in the figure. }
\end{center}
\end{figure}

In Fig. 9, we present the rates of SNe Ia in the Galaxy by adopting a
metallicity $Z=0.02$ and star-formation rate of $5\,M_{\rm \odot}{\rm
  yr}^{-1}$.  This work leads to a SN Ia rate of $\sim$$0.2\times
10^{-3}\,{\rm yr}^{-1}$ (thick curves in Fig. 9) based on the initial
parameter space in Fig. 8, which is lower than the observed estimate
of $\sim$$3\times 10^{-3}\ {\rm yr}^{-1}$ (e.g. Cappellaro \& Turatto
1997).  This indicates that the He star donor channel only contributes
to a small fraction of all SNe Ia ($\sim$7\%); some other mechanisms
or formation channels are therefore required to produce all SNe Ia
(see, e.g. Tout 2005; Wang \& Han 2012).  If we adopt the larger
initial contours for producing SNe Ia in Wang et al. (2009a), the SN
Ia rates will increase to $\sim$$0.3\times 10^{-3}\,{\rm yr}^{-1}$
(thin curves in Fig. 9), which is higher than the results in the
present work but not significantly so.  The reason is that WD binaries
with larger He donor masses, that also form SNe Ia in Wang et
al. (2009a), do not contribute significantly in the BPS simulations.
Accounting for off-centre carbon ignition changes the overall
  SN Ia rate of this channel but has no effect on the shape of the
  delay-time distribution.  Note that SNe Ia from this channel happen
systemically earlier in high metallicity environments (see Wang \& Han
2010b).

\section{Discussion}

In the classical DD model, SNe Ia arise from the merging of two CO WDs
that are brought together by gravitational wave radiation and have a
total mass above the Chandrasekhar limit (e.g. Webbink 1984; Iben \& Tutukov
1984; Yungelson \& Kuranov 2017).  However, a fundamental challenge to
this model is that off-centre carbon ignition due to a high
mass-accretion rate during the merger (or in the post-merger cooling
phase; Yoon, Podsiadlowski \& Rosswog 2007) is likely to convert the
CO WDs to ONe WDs through an inward-propagating carbon flame; ONe WDs
would collapse into neutron stars as core accretion continues
(e.g. Saio \& Nomoto 1985, 1998; Nomoto \& Iben 1985; Kawai, Saio \&
Nomoto 1987; Timmes, Woosley \& Taam 1994; Yoon et al. 2007; Shen et
al. 2012; Schwab, Quataert \& Bildsten 2015).  For the merging of two
CO WDs, it has been suggested that the critical mass-accretion rate
for off-centre carbon burning is close to $2\times
10^{-6}\,{M}_\odot\,\mbox{yr}^{-1}$ (e.g. Saio \& Nomoto 1985; Kawai,
Saio \& Nomoto 1987).  In this work, we also found that off-centre
carbon ignition will happen if the accretion rate is above a critical
value ($\sim$$2.05\times 10^{-6}\,{M}_\odot\,\mbox{yr}^{-1}$).  By
considering off-centre carbon burning, we found that the Galactic
SN Ia rate from the He star donor channel decreases from
  $\sim$$0.3\times 10^{-3}\,{\rm yr}^{-1}$ to $\sim$$0.2\times
10^{-3}\,{\rm yr}^{-1}$. On the other hand, off-centre carbon burning
contributes to the formation of neutron stars, which will increase the
rate of accretion induced collapse that needs to be studied in the
future.

In the present work, we used hot WDs to simulate He accretion
  onto WDs.  Chen et al. (2014) recently explored the effect of
  different cooling times and found that it affects the conditions
  at carbon ignition (see also Lesaffre et al. 2006; Brooks et
  al. 2016).  It may thus also affect the critical accretion rate
  separating centre and off-centre carbon ignition; a cold WD needs a
  thick He layer for off-centre carbon ignition.  Another limitation
  is the (necessary) assumption of spherical symmetry in the stellar
  evolution code. Any possible 3D effects on the ignition of carbon,
  especially for off-centre conditions, are therefore ignored. In
  reality, ignition will start at one location in the shell and
  not simultaneously over the entire shell. Most relevant for this
  study is the fact that the accreting WD will be rotating quite
  rapidly, which may change the conditions for off-centre carbon
  ignition and may also affect how carbon burning proceeds
  afterwards. These uncertainties should be explored in the future.

It has been suggested that HD 49798 (a hydrogen stripped subdwarf O6
star) with its X-ray pulsating companion (a massive WD) is a strong
progenitor candidate for a SN Ia (e.g. Wang \& Han 2010a; Mereghetti
et al. 2011; Liu et al. 2015).  Due to the rapid expansion of HD
49798's envelope, it will fill its Roche lobe in about $4
\times10^{4}$\,yr (see Wang \& Han 2010a).  In this binary, off-centre
carbon burning may happen when the WD grows in mass close to the Chandrasekhar
limit due to the high mass-transfer rate ($>$$2.05\times
10^{-6}\,{M}_\odot\,\mbox{yr}^{-1}$; see Fig.  2 of Wang \& Han
2010a). Thus, we speculate that the massive WD in this binary may
eventually form a neutron star and not produce a SN Ia.  In addition,
SN 2012Z may originate from the evolution of a hybrid CONe WD+He star
system (see Wang et al. 2014a).

For the He star donor channel, the mass donor would survive and should
be observable after the SN explosion (see Wang \& Han 2009).  The
identification of the surviving donors would support this progenitor
channel (e.g. Podsiadlowski 2010; Liu et al. 2012).  Wang \& Han (2009) studied the
properties of the surviving donors of SNe Ia based on the He star
donor channel and suggested that this channel is an alternative way
for the formation of hypervelocity He stars such as US 708 (see also
Justham et al. 2009; Geier et al. 2015). In order to study
the surviving donors of the He star donor channel, some hydrodynamical
simulations related to the impact of the SN explosion on the He donors
were performed (e.g. Pan, Ricker \& Taam 2010, 2013; Liu et al. 2013).
It can be expected that more hypervelocity He stars originating as
surviving donors of SNe Ia are discovered by some of the ongoing
surveys, e.g.  the Hyper-MUCHFUSS project (e.g. Geier et al.  2011,
2015; Tillich et al. 2011) and the LAMOST LEGUE survey (e.g. Deng et
al. 2012).

In this work, we calculated the long-term evolution of He-accreting
WDs with $\dot{M}_{\rm
  acc}>4\times10^{-8}\,{M}_\odot\,\mbox{yr}^{-1}$.  However, if
$\dot{M}_{\rm acc}$ is too low
(e.g. $<$$4\times10^{-8}\,{M}_\odot\,\mbox{yr}^{-1}$), the CO core
cannot grow in mass but a thick He-shell will develop on the surface
of the WD (e.g. Woosley, Taam \& Weaver 1986). Under these conditions,
a double-detonation may occur in a sub-Chandrasekhar WD if the mass of the
He-shell reaches a critical value (e.g. Nomoto 1982b; Iben \& Tutukov
1989; Livne 1990; H\"{o}flich \& Khokhlov 1996; Neunteufel, Yoon \&
Langer 2016).  Wang, Justham \& Han (2013) obtained the parameter
space in the double-detonation model for producing SNe Ia and
suggested that this model could explain the formation of type Iax SNe
(a sub-type of sub-luminous SNe Ia similar to SN 2002cx; e.g. Foley et
al. 2013).  It has been thought that CD$-$30$^{\circ}$\,11223 (a WD+He
star system) may form a SN Ia via the double-detonation model in its
future evolution (see Wang, Justham \& Han 2013; Geier et
al. 2013). Note, however, that current simulations of the
double-detonation model still fail to reproduce many of the main
properties of observed SNe Ia (e.g. Kromer et al. 2010).

The present work only involved single-shell (He-shell) burning on the
surface of a WD, in which the WD can grow in mass to the Chandrasekhar limit
through steady He-shell burning and finally produce a SN Ia.  In the
standard SD model, a WD can also obtain H-rich matter from its
non-degenerate donor, which leads to double-shell (H-/He-shell)
burning; the accumulated H-rich matter in this case is first burnt
into He and then converted into carbon and oxygen. However, it is
still difficult for the WD to grow in mass to the Chandrasekhar limit as stable H
and He burning require different mass growth rates (e.g. Idan, Shaviv
\& Shaviv 2013; Hillman et al. 2016).  This fundamental difficulty for
double-shell burning on the surface of the WD needs to be resolved in
future studies.

\section{Summary}
 By employing the MESA stellar evolution code, we carried out a series
 of simulations of He accretion onto CO WDs.  In each simulation, we
 calculated the evolution of the He-accreting WD for a sufficiently
 long time to determine their detailed properties at the time of
 carbon ignition.  We found that He-accreting WDs in the stable He
 burning region will experience either centre or off-centre carbon
 ignition when the WDs approach the Chandrasekhar limit.  When off-centre carbon
 burning is included, the Galactic rate of SNe Ia for the He star
 donor channel decreases from $\sim$$0.3\times 10^{-3}\,{\rm
     yr}^{-1}$ to $\sim$$0.2\times 10^{-3}\,{\rm yr}^{-1}$.
 Importantly, this work indicates that a WD can grow in mass to the Chandrasekhar
 limit through steady He accretion and eventually produce a SN Ia.  We
 also produce the chemical abundance profile of the WD at the time of
 the SN explosion, which can be used as initial input for SN Ia
 explosion models.  To set constraints on the He star donor channel,
 large samples of observed massive WD+He star systems and the
 surviving donors are needed. Finally, the process of H/He accretion
 onto WDs still needs further study.

\section*{Acknowledgments}
We acknowledge the anonymous referee for the valuable comments that helped us
to improve the paper.
This study is supported by the NSFC (Nos 11673059, 11390374  and 11521303),
the Yunnan Province (Nos 2013HA005 and 2013HB097),  and
the CAS (Nos QYZDB-SSW-SYS001 and KJZD-EW-M06-01).

\label{lastpage}

\begin{thebibliography}{}
\bibitem[Bours, Toonen \& Nelemans (2013)]{bou13}   Bours M. C. P., Toonen S.,  Nelemans G., 2013, A\&A, 552, A24
\bibitem[Brooks et al. (2015)]{bro15}                   Brooks J., Bildsten L., Marchant P., Paxton B., 2015, ApJ, 807, 74
\bibitem[Brooks et al. (2016)]{bro16}                   Brooks J., Bildsten L., Schwab J., Paxton B., 2016, ApJ, 821, 28
\bibitem[Cappellaro \& Turatto (1997)]{CT97}     Cappellaro E.,  Turatto M., 1997, in Ruiz-Lapuente P., Cannal R.,
Isern J., eds, Thermonuclear Supernovae, Kluwer, Dordrecht, P. 77
\bibitem[Chen \& Li (2009)]{che09}                     Chen W.-C., Li X.-D., 2009, ApJ, 702, 686
\bibitem[Chen, Han \& Meng (2014)]{che14}       Chen X.-F., Han Z., Meng X.-C., 2014, MNRAS, 438, 3358
\bibitem[Claeys et al. (2014)]{cla14}                     Claeys J. S. W., Pols O. R., Izzard R. G., Vink J.,  Verbunt F. W. M., 2014, A\&A, 563, A83
\bibitem[Cooper, Newman \&  Yan   (2009)]{cny09}   Cooper M. C., Newman J. A., Yan R., 2009, ApJ, 704, 687
\bibitem[Deng et al. (2012)]{den12}                       Deng L. et al, 2012, Res. Astron. Astrophys., 12, 735
\bibitem[Diehl et al. (2014)]{die14}                      Diehl R. et al., 2014, Science, 345, 1162
\bibitem[Eggleton (1973)]{egg73}                         Eggleton P. P., 1973, MNRAS, 163, 279
\bibitem[Foley et al. (2013)]{fol13}                       Foley R. J. et al., 2013, ApJ, 767, 57
\bibitem[Geier et al. (2011)]{gei11}                       Geier S. et al.,  2011, A\&A, 530, A28
\bibitem[Geier et al. (2013)]{gei13}                       Geier S. et al., 2013, A\&A, 554, A54
\bibitem[Geier et al. (2015)]{gei15}                       Geier S. et al., 2015, Science, 347, 1126
\bibitem[Geier et al. (2007)]{GEI07}                     Geier S., Nesslinger S., Heber U., Przybilla N., Napiwotzki R., Kudritzki R.-P., 2007, A\&A, 464, 299
\bibitem[Hachisu, Kato \&  Nomoto (1996)]{hac96}   Hachisu I., Kato M., Nomoto K., 1996, ApJ, 470, L97
\bibitem[Hachisu et al. (2012)]{hac12}                   Hachisu I., Kato M., Saio H.,  Nomoto K., 2012, ApJ, 744, 69
\bibitem[Han \& Podsiadlowski (2004)]{han04}     Han Z., Podsiadlowski Ph., 2004, MNRAS, 350, 1301
\bibitem[Han \& Podsiadlowski (2006)]{han06}     Han Z., Podsiadlowski Ph., 2006, MNRAS, 368, 1095
\bibitem[Han, Podsiadlowski \& Eggleton (1994)]{han94}  Han Z., Podsiadlowski P., Eggleton P. P., 1994, MNRAS, 270, 121
\bibitem[Hillebrandt et al. (2013)]{hill13}              Hillebrandt W., Kromer M., R\"{o}pke F. K.,  Ruiter A. J., 2013, FrPhy, 8, 116
\bibitem[Hillman et al. (2016)]{Hill06}                  Hillman Y., Prialnik D., Kovetz A., Shara M. M., 2016, ApJ, 819, 168
\bibitem[H\"{o}flich et al. (2013)]{hoe13}             H\"{o}flich P.,  Dragulin P., Mitchell J., Penney B., Sadler B., Diamond T.,
Gerardy C., 2013, Frontiers Phys., 8, 144
\bibitem[H\"{o}flich  \& Khokhlov (1996)]{hoe96}    H\"{o}flich P., Khokhlov A., 1996, ApJ, 457, 500
\bibitem[Howell (2011)]{how11}                            Howell D. A., 2011, Nature Communications, 2, 350
\bibitem[Hoyle \& Fowler (1960)]{hoy60}             Hoyle F., Fowler W. A., 1960, ApJ, 132, 565
\bibitem[Hurley, Tout \& Pols (2002)]{Hur02}       Hurley J. R., Tout C. A., Pols O. R., 2002, MNRAS, 329, 897
\bibitem[Iben \& Tutukov (1984)]{it84}                 Iben I., Tutukov A. V., 1984, ApJS, 54, 335
\bibitem[Iben \& Tutukov (1989)]{it89}                  Iben I., Tutukov A. V., 1989, ApJ, 342, 430
\bibitem[Idan, Shaviv \& Shaviv (2013)]{ida13}     Idan I., Shaviv N. J., Shaviv G., 2013, MNRAS, 433, 2884
\bibitem[Justham (2011)]{jus11}                             Justham S., 2011, ApJ, 730, L34
\bibitem[Justham et al. (2009)]{Jus09}                   Justham S., Wolf C., Podsiadlowski P., Han Z., 2009, A\&A, 493, 1081
\bibitem[Kato \& Hachisu (2004)]{kh04}             Kato M., Hachisu I., 2004, ApJ, 613, L129
\bibitem[Kato et al. (2008)]{kh08}                       Kato M., Hachisu I., Kiyota S., Saio H., 2008, ApJ, 684, 1366
\bibitem[Kawai, Saio \& Nomoto (1987)]{ksn87}  Kawai Y., Saio  H., Nomoto K., 1987, ApJ,  315, 229
\bibitem[Kromer et al. (2010)]{kro10}                 Kromer M., Sim S. A., Fink M., R\"{o}pke F. K., Seitenzahl I. R., Hillebrandt W., 2010, ApJ, 719, 1067
\bibitem[Langer et al. (2000)]{lan00}                 Langer N., Deutschmann A., Wellstein S., H\"{o}flich P., 2000, A\&A, 362, 1046
\bibitem[Lesaffre et al. (2006)]{les06}               Lesaffre P., Han Z., Tout C. A., Podsiadlowski Ph.,  Martin R. G., 2006, MNRAS, 368, 187
\bibitem[Li \& van den Heuvel (1997)]{li97}      Li X.-D., van den Heuvel E. P. J., 1997, A\&A, 322, L9
\bibitem[Liu et al. (2016)]{Liu16}                       Liu D.,Wang B., Podsiadlowski Ph., Han Z., 2016, MNRAS, 461, 3653
\bibitem[Liu et al. (2015)]{Liu15}                       Liu D., Zhou W., Wu C., Wang B., 2015, Res. Astron. Astrophys., 15, 1813
\bibitem[Liu et al. (2010)]{Liu10}                       Liu W., Chen W., Wang B., Han Z., 2010, A\&A, 523, A3
\bibitem[Liu et al. (2013)]{liu13}                        Liu Z. et al., 2013, ApJ, 774, 37
\bibitem[Liu et al. (2012)]{liu12}                        Liu Z., Pakmor R., R\"{o}pke F. K., Edelmann P., Wang B.,  Kromer M., Hillebrandt W., Han Z., 2012, A\&A, 548, A2
\bibitem[L\"{u} et al. (2009)]{lv09}                      L\"{u} G., Zhu C., Wang Z., Wang N., 2009, MNRAS, 396, 1086
\bibitem[Mannucci, Della Valle \& Panagia (2006)]{man06}    Mannucci F., Della Valle M., Panagia N., 2006, MNRAS, 370, 773
\bibitem[Maoz, Mannucci \& Nelemans (2014)]{mao14}   Maoz D., Mannucci F., Nelemans G., 2014, ARA\&A, 52, 107
\bibitem[Matteucci \& Greggio (1986)]{Mat86}  Matteucci F., Greggio L., 1986, A\&A, 154, 279
\bibitem[Maxted, Marsh \& North (2000)]{max00}            Maxted P. F. L., Marsh T. R., North R. C., 2000, MNRAS, 317, L41
\bibitem[McCully et al. (2014)]{mcc14}               McCully C. et al., 2014, Nature, 512, 54
\bibitem[Meng, Gao \& Han (2015)]{men15}       Meng X., Gao Y., Han Z., 2015, IJMPD, 24, 1530029
\bibitem[Meng \& Podsiadlowski (2017)]{Men17}   Meng X., Podsiadlowski Ph., 2017, MNRAS, 469, 4763
\bibitem[Meng \& Yang (2012)]{Men12}             Meng X., Yang W., 2012, A\&A, 543, A137
\bibitem[Mereghetti et al. (2011)]{mer11}            Mereghetti S. et al.,  2011, ApJ, 737, 51
\bibitem[Nelemans et al. (2001)]{nel01}              Nelemans G., Yungelson L. R., Portegies Zwart S. F., Verbunt F., 2001, A\&A, 365, 491
\bibitem[Neunteufel, Yoon \& Langer (2016)]{neu16}         Neunteufel P., Yoon S.-C., Langer N., 2016, A\&A, 589, A43
\bibitem[Nomoto (1982a)]{nom82a}                    Nomoto K., 1982a, ApJ, 253, 798
\bibitem[Nomoto (1982b)]{nom82b}                    Nomoto K., 1982b, ApJ, 257, 780
\bibitem[Nomoto \& Iben (1985)]{nom85}           Nomoto K., Iben I., 1985, ApJ, 297, 531
\bibitem[Nomoto, Iwamoto \& Kishimoto (1997)]{nom97}  Nomoto K., Iwamoto K., Kishimoto N., 1997, Science, 276, 1378
\bibitem[Pan, Ricker  \& Taam (2010)]{pan10}    Pan K.-C., Ricker P. M., Taam R. E., 2010, ApJ, 715, 78
\bibitem[Pan, Ricker  \& Taam (2013)]{pan13}    Pan K.-C., Ricker P. M., Taam R. E., 2013, ApJ, 773, 49
\bibitem[Paxton et al. (2011)]{pax11}                   Paxton B., Bildsten L., Dotter A., Herwig F., Lessafre P.,  Timmes F., 2011, ApJS, 192, 3
\bibitem[Paxton et al. (2013)]{pax13}                   Paxton B. et al., 2013, ApJS, 208, 4
\bibitem[Paxton et al. (2015)]{pax15}                   Paxton B. et al., 2015, ApJS, 220, 15
\bibitem[Piersanti, Tornamb\'{e} \&  Yungelson (2014)]{pier14}  Piersanti L., Tornamb\'{e} A., Yungelson L. R., 2014, MNRAS, 445, 3239
\bibitem[Piersanti, Tornamb\'{e} \&  Yungelson (2015)]{pier15}  Piersanti L., Tornamb\'{e} A., Yungelson L. R., 2015, MNRAS, 452, 2897
\bibitem[Podsiadlowski (2010)]{pod10}             Podsiadlowski Ph., 2010, Astron. Nachr., 331, 218
\bibitem[Podsiadlowski et al. (2008)]{pod08}     Podsiadlowski Ph., Mazzali P., Lesaffre P., Han Z., F\"{o}rster F., 2008, New Astron. Rev., 52, 381
\bibitem[Pols et al. (1998)]{pol98}                      Pols O. R., Schr\"{o}der K. P., Hurly J. R., Tout C. A., Eggleton P. P., 1998, MNRAS, 298, 525
\bibitem[Ruiter et al. (2014)]{Rui14}                    Ruiter A. J., Belczynski K.,  Sim S. A., Seitenzahl I. R., Kwiatkowski D., 2014, MNRAS, 440, L101
\bibitem[Ruiter et al. (2013)]{Rui13}                   Ruiter A. J. et al., 2013, MNRAS, 429, 1425
\bibitem[Ruiz-Lapuente (2014)]{rui14}               Ruiz-Lapuente P., 2014, New Astron. Rev., 62, 15
\bibitem[Saio \& Nomoto (1985)]{sn85}             Saio H., Nomoto K., 1985, A\&A, 150, L21
\bibitem[Saio \& Nomoto (1998)]{sn98}             Saio H., Nomoto K., 1998, ApJ, 500, 388
\bibitem[Schwab, Quataert \& Bildsten (2015)]{sch15}   Schwab J., Quataert E., Bildsten L., 2015, MNRAS, 453, 1910
\bibitem[Schwab, Quataert \& Kasen (2016)]{sch16}       Schwab J., Quataert E., Kasen D., 2016, MNRAS, 463, 3461
\bibitem[Shen et al. (2012)]{she12}                      Shen K. J., Bildsten L., Kasen D., Quataert E., 2012, ApJ, 748, 35
\bibitem[Thomson \& Chary (2011)]{tc11}         Thomson M. G., Chary R. R., 2011, ApJ, 731, 72
\bibitem[Tillich et al. (2011)]{til11}                    Tillich A. et al.,  2011, A\&A, 527, A137
\bibitem[Timmes, Woosley \& Taam (1994)]{tim94}    Timmes F. X., Woosley S. E., Taam R. E., 1994, ApJ, 420, 348
\bibitem[Toonen et al. (2014)]{ton14}                  Toonen S., Claeys J. S. W., Mennekens N.,  Ruiter A. J.,  2014, A\&A, 562, A14
\bibitem[Toonen et al. (2012)]{Too12}                Toonen S., Nelemans G., Portegies Zwart S., 2012, A\&A, 546, A70
\bibitem[Tout (2005)]{tou05}                               Tout C. A., 2005, ASPC, 330, 279
\bibitem[Vennes et al. (2012)]{ven}                     Vennes S., Kawka A., O'Toole S. J., N$\acute{\rm e}$meth P, Burton D., 2012, ApJ, 759, L25
\bibitem[Wang \& Han (2009)]{wh09}                Wang B., Han Z., 2009, A\&A, 508, L27
\bibitem[Wang \& Han (2010a)]{wh10a}             Wang B., Han Z., 2010a, Res. Astron. Astrophys., 10, 681
\bibitem[Wang \& Han (2010b)]{wh10b}             Wang B., Han Z., 2010b, A\&A, 515, A88
\bibitem[Wang \& Han (2012)]{wh12}                 Wang B., Han Z., 2012, New Astron. Rev., 56, 122
\bibitem[Wang, Justham \& Han (2013)]{wan13} Wang B., Justham S., Han Z., 2013, A\&A, 559, A94
\bibitem[Wang, Li \& Han (2010)]{wan10}          Wang B., Li X., Han Z., 2010, MNRAS, 401, 2729
\bibitem[Wang et al. (2015)]{wan15}                    Wang B., Li Y., Ma X.,  Liu D., Cui X., Han Z., 2015, A\&A, 584, A37
\bibitem[Wang et al. (2014a)]{wan14a}                Wang B., Meng X., Liu D., Liu Z., Han Z., 2014a, ApJL, 794, L28
\bibitem[Wang et al. (2014b)]{wan14b}                 Wang B., Justham S., Liu Z., Zhang J., Liu D., Han Z., 2014b, MNRAS, 445, 2340
\bibitem[Wang et al. (2009a)]{wan09a}                Wang B., Meng X., Chen X., Han Z., 2009a, MNRAS, 395, 847
\bibitem[Wang et al. (2009b)]{wan09b}                Wang B., Chen X., Meng X., Han Z., 2009b, ApJ, 701, 1540
\bibitem[Webbink (1984)]{web84}                     Webbink R. F., 1984, ApJ, 277, 355
\bibitem[Whelan \& Iben (1973)]{whe73}          Whelan J., Iben I., 1973, ApJ, 186, 1007
\bibitem[Woosley, Taam \& Weaver (1986)]{woo86} Woosley S. E., Taam R. E., Weaver T. A., 1986, ApJ, 301, 601
\bibitem[Woudt et al. (2009)]{wou09}                Woudt P. A. et al., 2009, ApJ, 706, 738
\bibitem[Wu et al. (2016)]{Wu16}                       Wu C., Liu D., Zhou W., Wang B., 2016, Res. Astron. Astrophys., 16, 160
\bibitem[Wu et al. (2017)]{Wu17}                       Wu C., Wang B., Liu D., Han Z., 2017, A\&A, 604, A31
\bibitem[Yoon \& Langer (2003)]{yoo03}           Yoon S.-C., Langer N., 2003, A\&A, 412, L53
\bibitem[Yoon \& Langer (2004)]{yoo04}           Yoon S.-C., Langer N., 2004, A\&A, 419, 623
\bibitem[Yoon, Podsiadlowski \& Rosswog (2007)]{yoo07} Yoon S.-C., Podsiadlowski Ph., Rosswog S., 2007, MNRAS, 380, 933
\bibitem[Yungelson \& Kuranov (2017)]{yun17}   Yungelson L. R., Kuranov A. G., 2017, MNRAS, 464, 1607
\bibitem[Ziegerer et al. (2017)]{zie17}                   Ziegerer E., Heber U., Geier S., Irrgang A., Kupfer T., F\"urst F., Schaffenroth J., 2017, A\&A, 601, A58

\end{thebibliography}
\end{document}